\documentclass[12pt]{iopart}
\usepackage{graphicx}
\usepackage{iopams}

\jl{6}

\renewcommand{\Re}{\mbox{Re}}
\renewcommand{\Im}{\mbox{Im}}

\begin{document}
\title[Black Hole Spectroscopy]{Black Hole Spectroscopy: Testing General
Relativity through Gravitational Wave Observations}
\author{
Olaf Dreyer,\dag
Bernard Kelly,\ddag
Badri Krishnan,\S
Lee Samuel Finn,$\|$\footnote[5]{To whom correspondence should be addressed.}
David Garrison,$^+$, and
Ramon Lopez-Aleman*}

\address{\dag\ Perimeter Institute of Theoretical Physics, 35 King
Street North, Waterloo, Ontario, N2J 2G9 Canada; E-mail:odreyer@perimeterinstitute.ca}

\address{\ddag\ Center for Gravitational Wave Physics, Center for Gravitational Physics
and Geometry, and Department of Physics, 
104 Davey Laboratory, University Park PA 16802, USA; E-mail: kelly@gravity.psu.edu}

\address{\S\ Max Planck Institut f\"ur
Gravitationsphysik, Am M\"{u}hlenberg 1, D-14476 Golm, Germany; E-mail: badkri@aei.mpg.de}

\address{$\|$\ Center for Gravitational Wave Physics, Center for Gravitational Physics
and Geometry, Department of Physics and Department of Astronomy and Astrophysics, 
104 Davey Laboratory, University Park PA 16802, USA; E-mail: lsfinn@psu.edu}

\address{$^+$\ University of Houston, Clear
Lake, 2700 Bay Area Bvd, Room 3531-2, Houston, TX 77058 USA; E-mail:garrison@cl.uh.edu}

\address{*\ Physical Sciences Department, University
of Puerto Rico, Rio Piedras Campus, Rio Piedras, Puerto Rico 00931; E-mail: rlopez@uprrp.edu}

\begin{abstract}
    
    Assuming that general relativity is the correct theory of gravity in the strong field limit, 
    can gravitational wave observations distinguish between black hole 
    and other compact object sources? Alternatively, can gravitational wave observations 
    provide a test of one of the fundamental predictions of general relativity? 
    Here we describe a definitive test of the
    hypothesis that observations of damped, sinusoidal gravitational waves
    originated from a black hole or, alternatively, that nature respects the 
    general relativistic no-hair theorem. For
    astrophysical black holes, which have a negligible charge-to-mass
    ratio, the black hole quasi-normal mode spectrum is characterized 
    entirely by the black hole
    mass and angular momentum and is unique to black holes. 
    In a different theory of gravity,
    or if the observed radiation arises from a different source (e.g., a neutron 
    star, strange matter or boson star), the spectrum will be inconsistent
    with that predicted for general relativistic black holes. We give a statistical 
    characterization of the
    consistency between the noisy observation and the theoretical
    predictions of general relativity, together with a numerical
    example.
\end{abstract}


\maketitle

\section{Introduction}

The formation of a black hole is the ultimate manifestation of 
strong field
gravity. During the late
stages in the aspherical formation of an astrophysical black hole 
the
gravitational waves emitted are dominated by a set of quasi-normal
modes (QNMs) \cite{cunningham78,cunningham79,cunningham80}:
exponentially damped sinusoids, whose frequency and damping times are
characteristic of the black hole's mass and angular momentum (And 
electric charge, as well; however, astrophysical black
holes, which are our interest here, have negligible charge-to-mass
ratio). Should we observe a QNM from a black hole and know also which 
particular
normal mode we are observing
we can determine, from the mode's frequency and damping time, the
black hole mass and angular momentum \cite{eche89,finn92a}.  

If, on the other hand, we observe several different QNMs from the 
same source and find
that they are inconsistent with the spectrum predicted by general
relativity in the sense that they \emph{cannot} be explained by a
single value of the mass and angular momentum we may infer that we
are not observing a black hole. Here we develop this observation 
into a experimental test of the existence of black holes or, alternatively, 
general relativity. 

Future observations by the Laser Interferometer Space Antenna 
(LISA) \cite{danzmann98a} offer us a different perspective on this question. 
LISA is expected to observe mergers of compact objects 
with masses in the range $10^{6}-10^{8}\,\mbox{M}_{\odot}$ \cite{finn00b}. 
In our present
understanding, these compact objects can only
be black holes. Observations by LISA of QNMs inconsistent with black holes 
would also be a test of the general relativistic no-hair theorem 
since an inconsistency in this mass range with black hole sources 
would indicate that physical scales other 
than mass and angular momentum were involved in the 
generation of the radiation. 

In either sense the test described here is of general relativity based on
gravitational wave observations. 
Eardley et al. \cite{eardley73a} proposed the first test of general
relativity using gravitational wave observations.  They investigated
the polarization modes of gravitational waves in various metric
theories of gravity and described how to identify the modes
experimentally and use those observations to identify the spin content
of dynamical gravity.  The first actual test of general relativity
relying on its prediction of the existence of gravitational waves was
made by Taylor and Weisberg \cite{taylor82a}.  They described how the
observed orbit and orbit decay of the Hulse-Taylor binary pulsar PSR
B1913$+$16 led to a strong consistency check on the predictions of
general relativity.  Finn \cite{finn85a} proposed a different test of
the spin content of dynamical gravity, based on the possibility of a
space-based detector in circumsolar orbit observing the induction-zone
field associated with solar oscillations.  Ryan \cite{ryan97a} has
outlined how observations of the gravitational radiation from capture
orbits of solar mass compact bodies about a supermassive black hole
may allow the determination of certain multipole moments of the
central hole, thereby testing the prediction of general relativity.
More recently, Will \cite{will98a} and Finn and Sutton \cite{finn01a,sutton02a} have 
described
tests of 
general relativity that bound the mass of the graviton, and 
Scharre and Will \cite{scharre02a} and Fairhurst
et al. \cite{ffww} have shown how gravitational wave observations of
pulsars may be used to bound the value of the Brans-Dicke coupling
constant. 

The preceding tests can be grouped into three different classes.  One set
of tests, which includes \cite{taylor82a,ryan97a,finn01a,scharre02a}, is based on
energy conservation arguments: the observed evolution of a system or
of the radiation from a system, is related to the energy loss expected
owing to the radiation.  A second class of tests, which includes
\cite{eardley73a,finn85a,ffww}, focuses on the observed polarization
modes of the field.  A third class, which includes \cite{will98a}, involves the 
frequency-dependent dispersion relationship associated with a massive graviton. 
The test described here 
is of a new class, based on the unique
character of the radiation spectrum associated with a disturbed black
hole.

QNMs appear as solutions to the equations describing
perturbations of a stationary black hole spacetime, subject to the
boundary conditions of no in-going radiation from infinity and no
up-coming radiation from the horizon.  The perturbation equations
describing Schwarzschild black holes were first described by Regge and
Wheeler \cite{regge57a} and Zerilli \cite{zerilli70a,zerilli71a}.  The
first QNM solutions to these equations were found by Press
\cite{press71a}. Teukolsky found the corresponding perturbation
equations for Kerr black holes \cite{teukolsky72a,teukolsky73a}, and,
with Press, first investigated their QNM solutions
\cite{press73a}. 

Damped sinusoidal motion is ubiquitous for systems approaching equilibrium
and one expects that collapse or coalescence will lead, in any theory
of gravity, to some form of QNM ringing.  If we observe a QNM spectrum
that is inconsistent with an isolated black hole, then there are two
possibilities.  On the one hand, general relativity may yet be
correct, but we are not observing an isolated black hole approaching
equilibrium\footnote{In fact, the uniqueness theorems have only been
proved for vacuum spacetimes and they are not true in the presence of
arbitrary matter fields or radiation. Nevertheless, it would be a
great surprise if the spacetime in the vicinity of a black hole is not
close to Kerr in some approximate sense.}.  Similarly, we may be
observing the radiation arising from a compact body that is not a black hole
--- e.g., a neutron star \cite{andersson95}, a boson star
\cite{seidel90c} or strange matter star \cite{alcock86}, whose QNM
spectrum will be determined by the properties and configuration of the
appropriate matter fields --- or a black hole carrying a previously
unknown macroscopic charge e.g., a dilaton field \cite{gibbons88}.  On
the other hand, general relativity may not be the correct theory of
gravity in the strong field limit.  Thus, while no single observation
may rule out general relativity, a set of observations, each of a different
source, \emph{none} of which is consistent with an isolated black
hole, could suggest the need to consider alternative theories of gravity in
the strong-field limit.

This paper is organized as follows.  In section \ref{idealqnmsec}
we briefly describe the QNMs of a Kerr black hole and explain how
the idealized observation of two or more modes in the
absence of noise enables us to extract the mass and angular
momentum of the black hole.  Real boats, of course, rock, and
section \ref{errorsec} generalizes the discussion to include
experimental errors and describes how one can use noisy
gravitational wave observations of QNMs to test general
relativity.  In section \ref{examplesec} we demonstrate, in a
model numerical simulation, the use of this method as applied to
LISA observations. Section \ref{sec:disc} investigates the range 
to which we can expect LISA to observe sources strong enough for 
this test to be applied. We conclude in section \ref{sec:conc} 
with a summary of our main results.

\section{Ideal observations}
\label{idealqnmsec}

\subsection{quasi-normal modes of Kerr black holes}

Following the aspherical collapse to a black hole, one
expects that the final black hole can be described as the perturbation
of a stationary Kerr hole. The dominant part of the gravitational
waves emitted as the black hole settles down can be described as a sum
over a countably infinite set of damped sinusoids, each characterized
by an amplitude, phase, frequency and damping time. (At still later
times, the radiation will be dominated by power-law tails arising from
the backscatter of radiation off the spacetime curvature in the
neighborhood of the black hole
\cite{cunningham78,cunningham79,cunningham80}; however, here we are
interested in the earlier, and higher amplitude, QNM ringing.) In this
sub-section we review those properties of the black hole QNM spectrum
that are important for our investigation; more detailed examinations
of the spectrum itself can be found in
\cite{leaver85a,onozowa97a,nollert99a,kokkotas99a}.

Gravitational wave detectors respond to a linear combination of the radiation
in the two polarization modes of the incident gravitational waves. In terms of 
the transvere traceless (TT) \cite{misner73a} gauge metric perturbation $h_{ij}$ 
the observable 
quantity $h(t)$ may be
written, for QNMs, in the form
\begin{equation}
    h(t)\simeq\Re\left[\sum_{l,m,n}
    A_{lmn}e^{-i\left(\omega_{n{\ell}m}t +\phi_{n{\ell}m}\right)}
    \right]
\end{equation}
where the summation indices characterize the particular mode, which 
is related to the angular dependence of the mode amplitude and phase on 
a sphere of constant (Boyer-Lindquist) radius about the black hole through
$\ell$ and $m$, and the ``harmonic'' through the index $n$:  
$\ell=2,3\ldots$, $|m|\leq \ell$ and $n=1,2\ldots$. For the
Schwarzschild geometry the symmetry is spherical, the appropriate
decomposition of the metric perturbation is given by the usual
spherical harmonics, and modes differing only in $m$ are
degenerate. For Kerr the symmetry is axisymmetric and the orthonormal
decomposition of the perturbation is by spheroidal harmonics
\protect\cite{teukolsky73a}. The amplitudes $A_{n{\ell}m}$ and phases
$\phi_{n{\ell}m}$ depend on the initial conditions and the relative
orientation of the detector and the source; however, the complex
frequency $\omega_{n{\ell}m}$ depends only on the intrinsic parameters
of the underlying black hole: i.e., its mass $M$ and angular momentum
$aM^{2}$. (We assume that the black hole carries no significant
electric charge.)

For fixed $a$ the complex frequency $\omega_{n{\ell}m}$ scales as
$M^{-1}$; thus, we define the dimensionless frequency
$\Omega_{n{\ell}m}$,
\begin{eqnarray}\label{momega}
    \Omega_{n{\ell}m} &:=& M \omega_{n{\ell}m}\\
    &:=& \left( 2 \pi F_{n{\ell}m} + {i\over T_{n{\ell}m}}\right)
\label{eqn:defOmega}
\end{eqnarray}
where $F_{n{\ell}m}$ and $T_{n{\ell}m}$ are the real dimensionless
frequency and damping time of the modes.  The corresponding
physical frequency $f_{n{\ell}m}$ and damping time
$\tau_{n{\ell}m}$ are given by
\begin{equation}
\omega_{n\ell m} = 2\pi f_{n\ell m} + i/\tau_{n\ell m} = 2\pi F_{n\ell m}/M + i/(MT_{n\ell m}).
    \label{eqn:ftau}
\end{equation}
(We use geometrical units with $G=1$ and $c=1$.) The dimensionless
$\Omega_{n{\ell}m}$ (or $F_{n{\ell}m}$ and $T_{n{\ell}m}$) depend
\textit{only} on the also dimensionless black hole angular
momentum parameter $a$.  Figure \ref{fig:Omega} shows
$\Omega_{n{\ell}m}$ as a function of $a$ for $n=1,2$ and
$\ell=2,3$, and $|m|\leq\ell$.

\subsection{From quasi-normal modes to testing relativity}
\label{sec:idealTest}

If we observe only one mode, characterized by its complex frequency
$\omega$ (cf. eq. \ref{eqn:ftau}), what can we say about the
underlying black hole?

\begin{figure}
  \begin{center}
  \includegraphics[height=8cm]{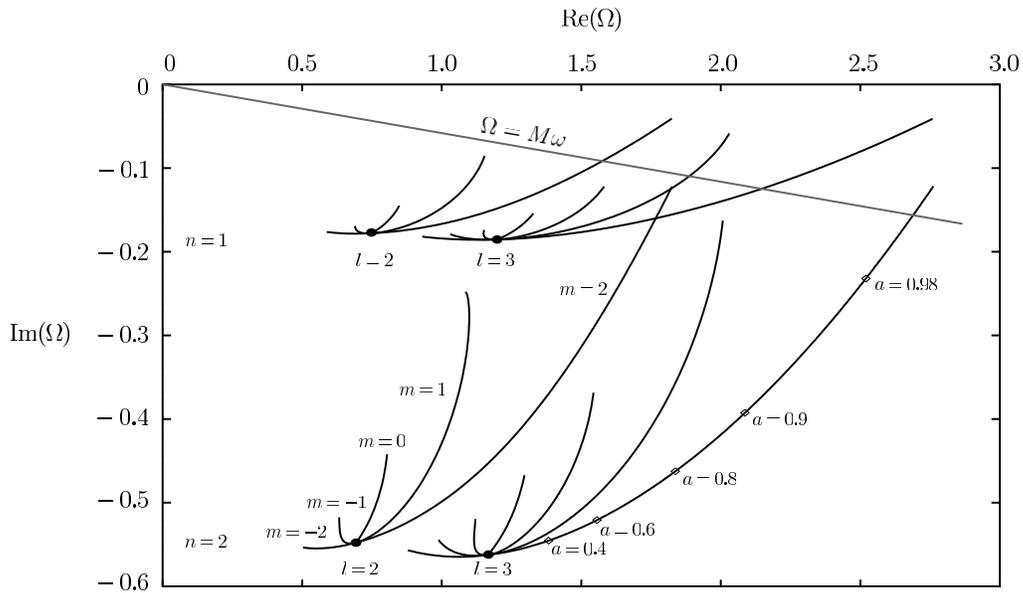}
  \caption{The dimensionless, complex QNM frequencies
  $\Omega_{n{\ell}m}$ for rotating, uncharged black holes.  Each
  family of curves corresponds to one $n{\ell}$ pair, and each branch
  to a possible value of $m$. The large black dot at the base of each
  family is the Schwarzschild ($a=0$) limit, where the frequencies are
  degenerate in $m$. This degeneracy is broken for $a\neq0$, and the
  curves emanating from the dots give the QNM frequencies for Kerr
  black holes as a function of positive $a$ for different $m$. In this
  figure $a$ ranges from $0$ to $0.9958$, with the small diamonds
  marking the QNM frequencies for $a=0.4, 0.6, 0.8, 0.9,$ and $0.98$.
  On this figure, an observation, corresponding to a (complex) frequency
  $\omega$, is represented
  by the line $\Omega=M\omega$, parameterized by the (unknown) 
  black hole mass
  $M$.  Each intersection of this line with a QNM curve in
  dimensionless $\Omega$ represents a candidate $n{\ell}m$, $M$ and
  $a$ for the mode.}\label{fig:Omega}
  \end{center}
\end{figure}

Corresponding to the observed $\omega$ is the line $\Omega = M\omega$,
$M \in \mathbb{R}_{\ge 0}$, in the dimensionless $\Omega$ plane (cf.\ eq.\
\ref{momega}). Such a line is shown in figure \ref{fig:Omega}.  This
line will intersect some subset of the family of $\Omega_{n{\ell}m}$
curves, characteristic of black hole normal modes.  Each intersection
corresponds to a black hole mass $M$, angular momentum
parameter $a$, and mode $n{\ell}m$ consistent with the observed $\omega$.  
Knowing only $f$ and $\tau$,
then, we cannot uniquely identify the black hole mass and angular momentum, 
but
we can reduce the possibilities to a (possibly countably infinite) set
of $(a,M)$ pairs.  If we knew $n{\ell}m$ as well, we would know $a$
and $M$ exactly.

\begin{figure}
  \begin{center}
      \includegraphics[height=6cm]{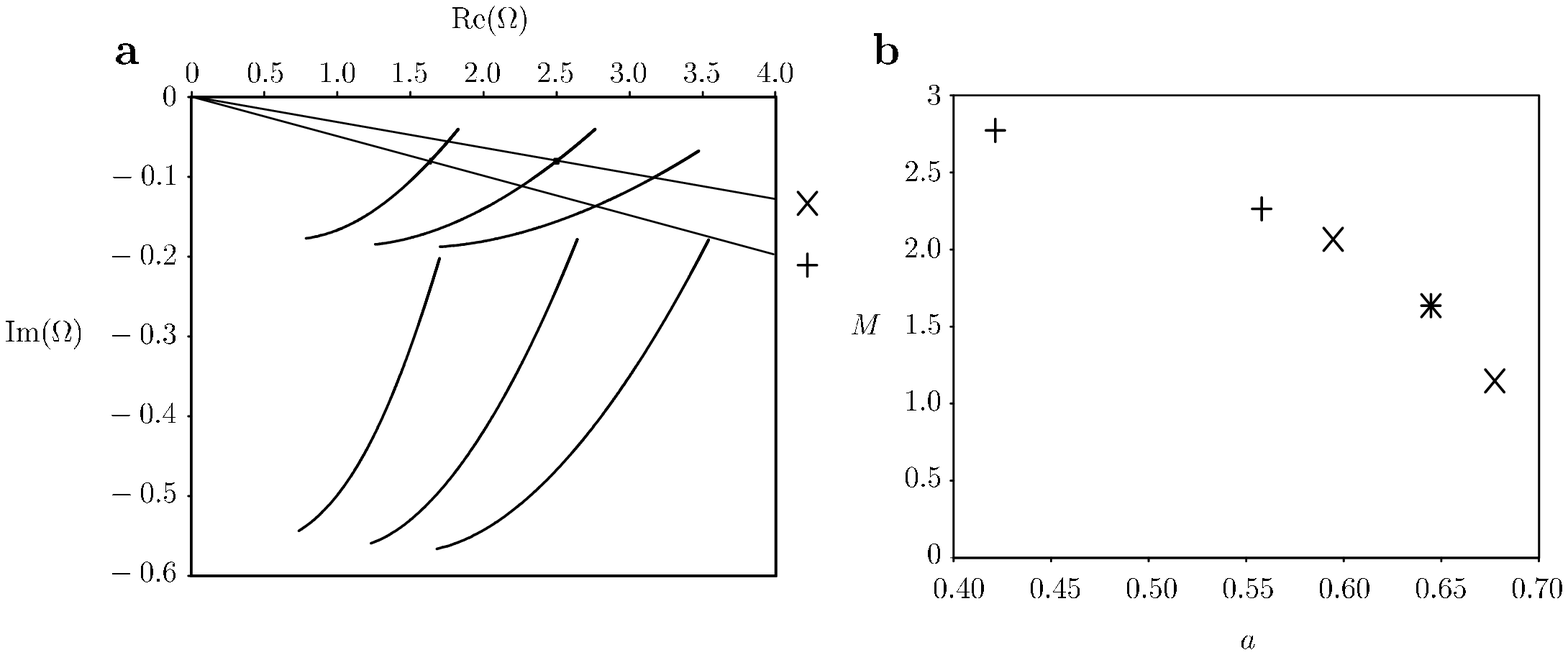}
      \caption{Here we show, in schematic form, several
      $\Omega_{n{\ell}m}(a)$ curves and their intersection with the
      lines $M\omega_i$, $M>0, i=1,2$, corresponding to two
      observed modes. These two lines we denote by $+$ and $\times$
      respectively.  (\textbf{b}) The candidate $(a,M)$ pairs
      determined in figure (a) are plotted here in the
      $(a,M)$--plane. The pairs belonging to $\omega_1$ are denoted by
      $+$, the ones belonging to $\omega_2$ by $\times$. There is only
      one candidate $(a,M)$ consistent with both observations --- indicated 
      by the overlapping $+$ and $\times$ --- and
      this is the actual mass and angular momentum of the underlying
      black hole.}\label{fig:ideal}
  \end{center}
\end{figure}

Now suppose that we observe two modes from the same black hole,
each characterized by its own frequency and damping time.  Figure
\ref{fig:ideal}a shows, in schematic form, the line $M\omega$ for each of
the two modes (denoted $+$ and $\times$) and their intersection
with several different $\Omega_{n{\ell}m}$ curves in the complex
$\Omega$ plane.  Corresponding to each mode is a set of candidate
$(a,M)$ pairs that may describe the underlying black hole.  Each
candidate mass and angular momentum parameter is a point in the
$(a,M)$ plane, as shown in figure \ref{fig:ideal}b.  With two or
more modes, there must be at least one common candidate mass and
angular momentum.

This is, in essence, our proposal for a test: interpreting the
observation of several normal modes $\omega_k$, $k\geq2$, 
as arriving from a single, general
relativistic black hole, we conclude that general relativity is
self-consistent if the observed $\omega_k$ are consistent with 
at least one black hole $(a,M)$. If no such $(a,M)$ exists for the observed
$\omega_k$ either
we have observed something other than an isolated black hole or we
have a contradiction with the predictions of the theory. 

(As an aside, it is possible (though unlikely) that we get more than
one value of $(a,M)$ consistent with the observed frequencies. This
can happen if we have two mode pairs $(n_1,\ell_1,m_1;n_2,\ell_2,m_2)$
and
$(\tilde{n}_1,\tilde{\ell}_1,\tilde{m}_1;\tilde{n}_2,\tilde{\ell}_2,\tilde{m}_2)$
which give rise to the same frequency $\omega$.  In this case the
observations would still be consistent with general relativity though
we could not use that observation to measure $M$ and $a$.  The
important point of our test is the existence of at least one $(a,M)$
pair consistent with the observations.)

Noise and other experimental realities ensure that there will be
no \emph{exact} agreement between the observed $\omega_k$ and
a general relativistic black hole even if general relativity is
correct. The challenge, then, in developing a practical test is to
determine when the differences between the candidate $(a,M)$ pairs
associated with the different observed modes are so great as to be
statistically inconsistent with general relativity.  In the next
section we face this challenge.

\section{A test of relativity}
\label{errorsec}

\subsection{A reformulation of the test}\label{sec:reformulate}

Before we discuss the role that noise plays in our analysis it is
helpful to reformulate the test described in section
\ref{sec:idealTest} and figure \ref{fig:ideal}.  Consider an
ordered $N$-tuple of QNMs,
\begin{equation}
    {\cal Q} := \left\{\left(n_{k},\ell_{k},m_{k}\right):
    k = 1\ldots N\right\}.
\end{equation}
Each $\cal Q$ may be regarded as a function that maps $M$ and $a$ to a
set of observable frequencies
\begin{equation}
    {\cal Q}(a,M) :=
    \left\{M^{-1}\Omega_{n_{k}\ell_{k}m_{k}}(a): k = 1\ldots N\right\}.
\end{equation}
Each $N$-tuple $\cal Q$ thus describes a two dimensional surface in the
$(2N+2)$-dimensional space ${\cal S}$,
\begin{equation}
    {\cal S} := (a,M,\omega_{1},\ldots,\omega_{N}),
\end{equation}
with different $N$-tuples corresponding to different sets of $N$
modes. (In section \ref{sec:criterion} we will understand the $\omega_k$ 
to represent observed QNM frequencies and damping times.)

\begin{figure}
  \begin{center}
  \includegraphics[height=6cm]{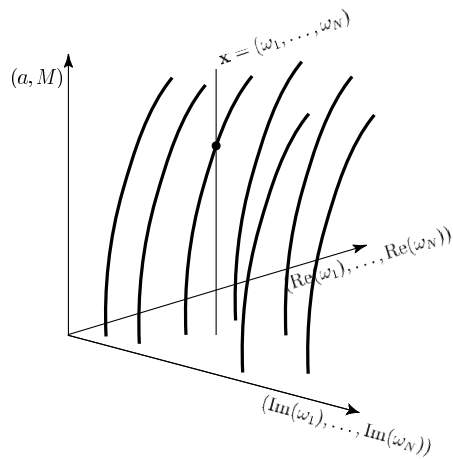}
  \caption{A reformulation of the consistency criterion. A set of
  quasi-normal modes ${\cal Q}=\{(n_k,l_k,m_k) : k=1,\ldots,N\}$ 
  corresponds to a surface in the $(2N + 2)$-dimensional
  space depicted in this figure. A measurement $\boldsymbol{\omega} =
  (\omega_1,\ldots,\omega_N)$ is consistent with general
  relativity if the constant surface that is obtained by ranging over
  all $(a,M)$ while keeping the frequencies $\boldsymbol{\omega}$ fixed
  intersects at least one of the surfaces corresponding to one of the
  sets $\cal Q$.  This intersection is indicated in this figure by a
  dot.}\label{idealsnakes}
  \end{center}
\end{figure}

An observation $\boldsymbol{\omega}$ consists of an $N$-tuple
\begin{equation}
    \boldsymbol{\omega} := \left(\omega_{1},\ldots, \omega_{N}\right).
\end{equation}
The observation $\boldsymbol{\omega}$ also corresponds to a surface in ${\cal
S}$.  The observation is consistent with a black hole if the surface
of constant $\boldsymbol{\omega}$ intersects one of the surfaces $\cal Q$.
Figure \ref{idealsnakes} shows a low-dimensional projection of such an
observation $\boldsymbol{\omega}$ together with several surfaces (which appear
as curves) for different $N$-tuples $\cal Q$. A moment's consideration
should convince one that this new criterion is equivalent to the
criterion formulated above in section \ref{sec:idealTest}.

In practice the situation is less ideal: noise distorts our
observation, so that --- even if we are observing black hole QNMs
--- the measured $\boldsymbol{\omega}$ will not intersect a curve 
${\cal Q}$. In the
remainder of this section we describe how our ideal test is made
practical and meaningful for real observations.

\subsection{Confidence intervals and testing general relativity}

In a frequentist analysis, the observation, the sampling distribution,
an ordering principle and a probability combine to determine a
confidence interval.  In this section we use this construction to form
a confidence region in the $(a,M)$-plane, given a noisy observation
$\boldsymbol{\omega}$.

\begin{figure}
  \begin{center}
  \includegraphics[height=6cm]{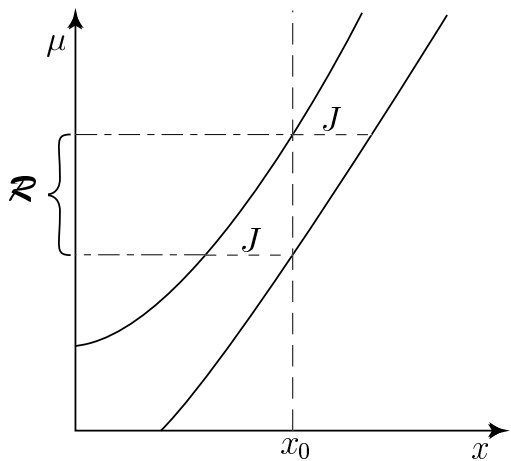}
  \caption{The construction of classical confidence intervals. A
  sampling distribution $P(x\vert\mu)$, an ordering principle, and
  a probability $p$ are needed to construct a confidence
  interval. The ordering principle is used to find the intervals
  $J(\mu)$ such that $\int_{J(\mu)} dx\; P(x\vert\mu) = p$. The
  classical confidence interval $\mathcal{R}$ is then given by the set of $\mu$
  for which $J(\mu)$ contains the measured value $x_0$.}\label{confidence}
  \end{center}
\end{figure}

We begin by reviewing the construction of a classical confidence
interval for the one-dimensional case following \cite{neyman37}
(alternatively, see e.g.\cite{stuart94a}).  We suppose that we make
measurements of a random variable $x$ from which a quantity $\mu$ is
determined. The sampling distribution $P(x|\mu)$ is the probability of
making the observation $x$ given a particular $\mu$.  Formally, an
\emph{ordering principle} is a function $R(x|\mu)$, which we use to
identify a sub-interval $J$ of $x$ according to
\begin{equation}
    J(\mu|r) := \left\{x: R(x|\mu)>r)\right\}.
\end{equation}
The parameter $r$ is chosen such that the region $J(\mu|r)$ encloses a
fixed probability $p$:
\begin{equation}
   \int_{J(\mu|r)}P(x|\mu)\,dx = p\,\, .
\end{equation}
Given an observation $x_0$, the probability-$p$ \emph{confidence
interval} is the range of $\mu$ for which $J(\mu|r(p))$ includes $x_0$
as shown in figure \ref{confidence}.  In an actual experiment, the
choice of the value of the parameter $p$ is made by the
experimentalist. Typical choices are 90\%, 95\% or 99\%. 

The choice of the ordering principle $R(x|\mu)$ is a 
key ingredient in the construction of confidence intervals. Different
choices will lead to different confidence intervals for the same
observation: for example, one choice of ordering principle
will always determine intervals of the form $(-\infty ,x)$, while another
choice will always determine intervals of the form
$(x,\infty )$. Neither choice is \emph{a priori} right or wrong.  
Here we will choose $R(x|\mu)=P(x|\mu)$
so that the intervals are given by level surfaces of the distribution 
$P(x|\mu)$. The main advantage of this ordering principle is that it is
simple and it works in any dimension. Consider, for example, a two-component
observation depending on one parameter $a$.  There is, as before, a sampling
distribution $P(x,y|a)$ and an ordering principle $R(x,y|a)=P(x,y|a)$.
Confidence intervals can be defined in the same way as in the case
of a one-dimensional observation; the interval $J$ is now a
two-dimensional region. Since this system is over-determined --- we are now 
trying to determine \emph{one} parameter $a$ by
measuring \emph{two} quantities $x$ and $y$ --- the 
measured $x$ and $y$ will have to satisfy additional
constraints in order to give a non-vanishing confidence
region.  This is in fact precisely what happens in the black
hole quasi-normal mode problem: any single measurement of $\omega$
can be explained by \emph{some} $(a,M)$, but a measurement of
two or more $\omega$ can be simultaneously consistent with at least 
one $(a,M)$ pair only if the no-hair theorem is true and the modes arise 
from a single black hole. 

We can now describe our test of relativity.  Note that not all
observations $\boldsymbol{\omega}$ will lead to a non-empty confidence
interval: i.e., for some $\boldsymbol{\omega}$ there will be no $(a,M)$
consistent with the observation. We frame our test in terms of the
confidence interval we construct: if we make an observation
$\boldsymbol{\omega}$ for which there is no probability $p$ confidence
interval, then we say that the observed normal modes are inconsistent
with an isolated black hole with confidence $p$.  Conversely, if there
does exist a probability $p$ confidence interval, then we have
verified that general relativity is self-consistent at this confidence
level.

Finally, we should point out an aesthetic flaw of our choice of 
ordering principle. The function  
$P(x|\mu)$ is a density and, therefore, not
invariant under a reparameterization of $x$. If we were to use a new parameter
$x^\prime=f(x)$ for some smooth monotonic function $f$, the confidence
region obtained for $\mu$ using a measurement of $x$ may not coincide
with the region obtained using a measurement of $x^\prime$.  In the
one-dimensional case, there exists another ordering principle based on
the likelihood ratio \cite{feldman98a} which is reparameterization
invariant; however, we have not been able to generalize this to higher
dimensions.  While aesthetically displeasing, there is nothing wrong 
with the choice we have made, which is natural given the physical 
association of the parameters $M$ and $a$ with the black hole mass and 
angular momentum. 

\subsection{Generalization to Quasi-normal Modes}
\label{sec:criterion}

The generalization to QNM observations is straightforward.  Each
observation consists of $N$ complex QNM frequencies $\omega_{k}$
and associated amplitude signal-to-noise ratios $\rho_{k}$, which
characterize both the amplitude of the signal at that frequency
and the uncertainty in the determination of $\omega_{k}$ (cf.\
\cite{finn92a}):
\begin{eqnarray}
    \boldsymbol{\omega} &:=& \left(\omega_1,\ldots,\omega_N\right)\\
    \boldsymbol{\rho} &:=& \left(\rho_1,\ldots,\rho_N\right).
\end{eqnarray}
For definiteness suppose that $\omega_{k}$ and $\rho_{k}$ are
identified via maximum likelihood techniques \cite{finn92a}.  There is
a minimum signal-to-noise associated with each mode, which is set by
the requirement that the observation must identify $N$ modes. 

Observations $\boldsymbol{\omega}$ corresponding to a black hole characterized 
by $(a,M)$ and signal-to-noise $\rho_k$ are distributed according to the 
sampling distribution
\begin{equation}
P(\boldsymbol{\omega}|a,M,{\cal Q},\boldsymbol{\rho}) := \left(\begin{array}{l}
\mbox{Probability of making observation}\\
\mbox{$\boldsymbol{\omega}$ given the actual $N$-tuple $\cal Q$}\\
\mbox{and signals-to-noise $\boldsymbol{\rho}$.}
\end{array}\right).
\end{equation}
In general the sampling distribution depends upon the nature of the detector noise
and the analysis procedure that identifies the modes $\omega_k$.
For large signal-to-noise ratios it will generally reduce 
to a multivariate Gaussian in $\Re(\omega_k)$ and $\Im(\omega_k)$ and for
smaller signal-to-noise ratios it can be
determined via simulation. 

\begin{figure}
  \begin{center}
      \includegraphics[height=10cm]{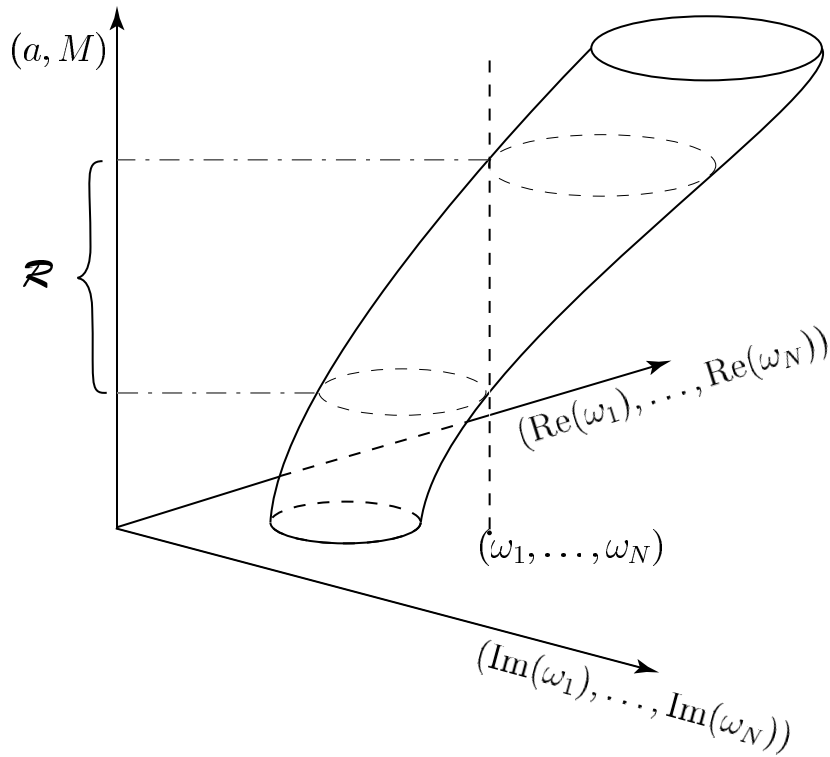}
      \caption{The construction of classical confidence intervals
      generalized to higher dimensions. Given a sampling
      distribution $P$, an ordering principle, and a probability
      $p$ one can construct classical confidence regions $R$ just
      as in the one-dimensional case. The difference here 
      is that we are now trying to determine a
      small number of parameters $(a,M)$ from a larger number of
      observations $\boldsymbol{\omega} = (\omega_1,\ldots,\omega_N)$. There
      are thus additional consistency conditions that need to be satisfied
      to obtain a non-empty confidence region $R$. This is the
      basis of our proposed test.}\label{snake}
  \end{center}
\end{figure}

Now consider the region of the space $\cal S$ (cf.\ section\ \ref{sec:reformulate})
defined by 
\begin{equation}\label{eq:regionA}
P(\boldsymbol{\omega}|a,M,{\cal Q},\boldsymbol{\rho})>p_0
\end{equation}
with $p_0$ such that 
\begin{equation}\label{eq:regionB}
p = \int_{P(\boldsymbol{\omega}|a,M,{\cal Q},\boldsymbol{\rho})>p_0}
P(\boldsymbol{\omega}|a,M,{\cal Q},\boldsymbol{\rho})d^{2N}\omega 
\end{equation}
for a fixed $p$. We say that the observation $\boldsymbol{\omega}$ is consistent 
with a black hole if the actual observation $\boldsymbol{\omega}$ 
is included in this region for some $(a,M)$. Figure \ref{snake} 
illustrates the comparison of an observation with the
region defined by equations (\ref{eq:regionA}--\ref{eq:regionB}). It remains only 
to specify $p$.  

To help in specifying $p$ it is useful to examine more closely its meaning. Suppose 
we have chosen a value of $p$. That value of $p$ determines a confidence
region. Now consider an ensemble of identical detectors, each observing 
simultaneously the same black hole event and its corresponding QNMs. The
fraction of these observations that does not intersect the confidence region is
the \emph{false alarm probability} $\alpha(p)$, so-called because it is the 
probability that an observation will be falsely deemed to be inconsistent with
a black hole. The probability $\alpha$ is a monotonic function of $p$; therefore, we 
can specify $\alpha$ in lieu of $p$. 
For observations whose characteristic frequency corresponds to 
masses greater than neutron star masses, which we are confident originate with 
black holes, we propose setting $p$ so that 
$\alpha(p)$ --- the probability of falsely rejecting the hypothesis that we have 
in fact observed a black hole --- is small (e.g., less than 1\%). In other words, the 
standard of evidence for declaring that we have discovered ``new physics'' should 
be high. 

The false alarm probability function $\alpha(p)$ will depend on the signal strength, 
as characterized by the signal-to-noise ratios; consequently, it will need to be 
determined on an observation-by-observation basis. Thus the
calculation of $\alpha(p)$ by a Monte Carlo simulation is the final
ingredient we need. In the next section we demonstrate the test
through a numerical example where we calculate $\alpha(p)$.

\section{A numerical example}
\label{examplesec}

In the previous section we described a general procedure for
testing general relativity by observing QNMs.  In this section 
we explore its effectiveness numerically through a set of simulated 
observations drawn from a hypothetical black hole population population
inspired by potential LISA observations, 
and a hypothetical population of non-black hole compact object sources, or NBHs. 
(We say ``inspired'' because, in fact, 
for the purpose of this analysis the observations are characterized
entirely by the dimensionless signal-to-noise ratio and mode quality
factor, with the dimensioned mode frequency simply setting a scale.
Thus, the conclusions we reach are as valid for LISA observations 
as they are for observations at the same signal-to-noise with 
ground-based detectors.)

For the BH observations we 
find the relationship between the false alarm probability $\alpha$ and 
the probability $p$ that appears in equation (\ref{eq:regionB}). For the NBH
observations there are no ``false alarms'': every
observation is of something not a black hole. Instead, there are
\emph{false dismissals}: observations that we mistakenly classify as 
consistent with a black hole. The probability of a false dismissal, denoted
$\beta$, depends on the choice of $p$ or, alternatively, the choice of
false alarm probability $\alpha(p)$ that we make for the purpose of 
defining the test. (The false dismissal probability depends
also on how the spectrum of BHs and NBHs differ.) The smaller 
the false dismissal probability the more sensitive the test is to discovering
``new physics'' or identifying non-black hole sources. For the NBH observations 
we evaluate the false dismissal probability 
as a function of the false alarm probability. 

\subsection{Simulating black hole QNM observations}
\label{sec:simObs}

For definiteness we focus on observations of two QNMs. For the purpose of illustration
we consider black hole masses and angular momenta consistent with potential 
observations by the LISA detector \cite{finn00b}. 
We first draw an $(a,M)$ pair from the distribution 
\begin{eqnarray}
P(a,M) &=& P(a)P(M)\\
P(a) &\propto& \left\{
\begin{array}{ll}
1&\mbox{for $a\in[0,0.986)$}\\
0&\mbox{otherwise}
\end{array}
\right.\\
P(M) &\propto& \left\{
\begin{array}{ll}
M^{-1}&\mbox{for $M\in(2.5\times10^5M_\odot,4.5\times10^8M_\odot)$}\\
0&\mbox{otherwise}
\end{array}
\right.\label{eq:distrM}
\end{eqnarray}
The range of $M$ is determined by the frequency band where LISA
is expected to be most sensitive; the range of $a$ is determined by 
the maximum angular momentum expected of a black hole spun-up
by thin-disk accretion \cite{thorne74a}. 

Corresponding to each $(a,M)$ pair we choose the QNMs corresponding to 
$(n=1, \ell = 2, m = 2)$ and $(n = 1, \ell = 4, m = 4)$. We assign each mode the
same signal-to-noise ratio, which we treat here as 
sufficiently large that the errors associated with the measurements are
normally distributed with covariance matrix $C_{ij}$ equal to the inverse 
of the \emph{Fisher information matrix} $I_{ij}$ (see e.g.
\cite{stuart94a}) as given in \cite[equation 4.14]{finn92a}.
This is in fact a mathematical lower bound --- the 
\emph{Cramer-Rao bound}  --- 
on the covariance matrix. We draw from this error distribution errors
in the frequencies and damping times that we add to the ``real''
frequencies and damping times to determine the simulated observations:
noisy QNM frequencies and damping times. 

Given this pair of QNM frequencies and damping times with errors we 
ask whether the two modes are in fact observationally distinguishable: if 
the frequencies and damping times are not sufficiently different then no
real observation would ever result in the given pair. For instance, the 
five 
$(n=1,l=2)$ modes are degenerate at $a=0$; consequently, no
matter how large the
signal-to-noise ratio, if 
$a$ is sufficiently small it is impossible to 
resolve these five modes observationally. 

To decide whether the two modes we are investigating are observationally 
distinguishable we invoke  a ``resolvability
criterion'': denoting the frequencies (damping times) 
of the two modes as $f_1$,
$f_2$ ($\tau_1$, $\tau_2$) we say that the two modes are distinguishable 
if
\begin{equation}
 | f_1 - f_2 | > \frac{1}{\min( \tau_1  , \tau_2 )}\, .
\end{equation}
We discard any mode pair that does not satisfy this criterion.

The result is an \emph{observation,}
which consists of a pair of signal-to-noise ratios and associated frequencies
and damping times. (The observation does \emph{not} include
knowledge of black hole mass or angular momentum, or the $n\ell m$ 
associated with the frequencies or damping times.)

\subsection{False alarm probability $\alpha$}\label{sec:alpha}
For each simulated observation $\boldsymbol{\omega}$, constructed as 
described in section 
\ref{sec:simObs} 
we evaluate the smallest probability $p = p_{\min}$ such that 
equations (\ref{eq:regionA}) and (\ref{eq:regionB}) describe a region $\cal S$ 
that covers 
$\boldsymbol{\omega}$ for some $(a,M)$. 
The false dismissal fraction $\alpha(p)$ is the fraction of $p_{\min}$ determinations
that are greater than $p$: i.e., the fraction of BH observations that we would reject as 
originating 
from a black hole for threshold $p$. 

\begin{figure}
  \begin{center}
  \includegraphics[height=10cm]{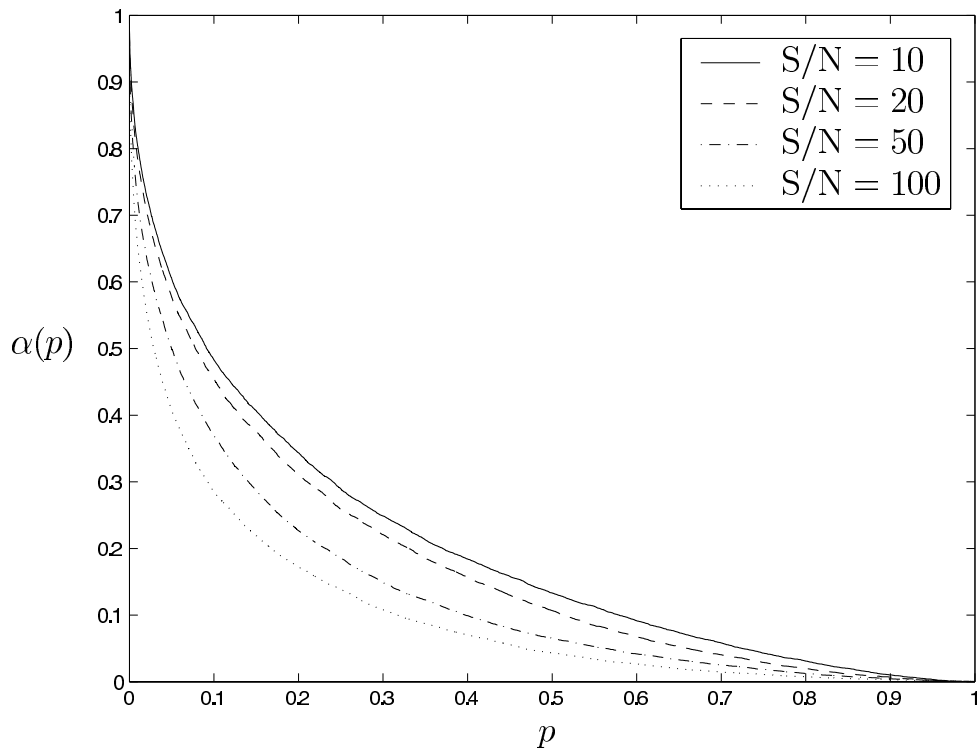}
  \caption{False alarm probability $\alpha$ as a function of the 
  probability $p$ appearing in equation (\ref{eq:regionB}). 
  A false alarm is a misidentification of a QNM pair as arising from 
  something other than a general relativistic black hole.}\label{alpha}
  \end{center}
\end{figure}

Ideally, in evaluating $p$ we would
consider every
possible $n{\ell}m$
for each $\omega_k$. In practice, we consider only a finite subset of low-order 
(in both $n$ and $\ell$) modes, corresponding to our expectation that
these are the 
modes most likely to be excited to large amplitude. In our simulations we 
considered only modes corresponding to $(n=1,\ell =2,m=0)$, 
$(n=1,\ell =2,m=2)$,
$(n=1,\ell =3,m=3)$ and $(n=1,\ell =4,m=4)$. Since for these simulations 
we observed
two distinguishable QNMs there were twelve possible ordered 
pairs of modes. 
Figure \ref{alpha} shows $\alpha$ as a function of $p$ for four
different signal-to-noise ratios. Each $\alpha(p)$ curve is constructed from 
$10^4$ simulated observations with that amplitude-squared signal-to-noise 
in each mode. 

\subsection{False dismissal probability calculation}
\label{sec:beta}

Complementary to $\alpha$, the probability that we incorrectly decide we have 
observed QNMs from something other than a black hole, is the probability that
we falsely conclude we have observed QNMs from a black hole. This probability 
is referred to as the false dismissal probability and commonly denoted $\beta$. 

\begin{figure}
  \begin{center}
  \includegraphics[height=10cm]{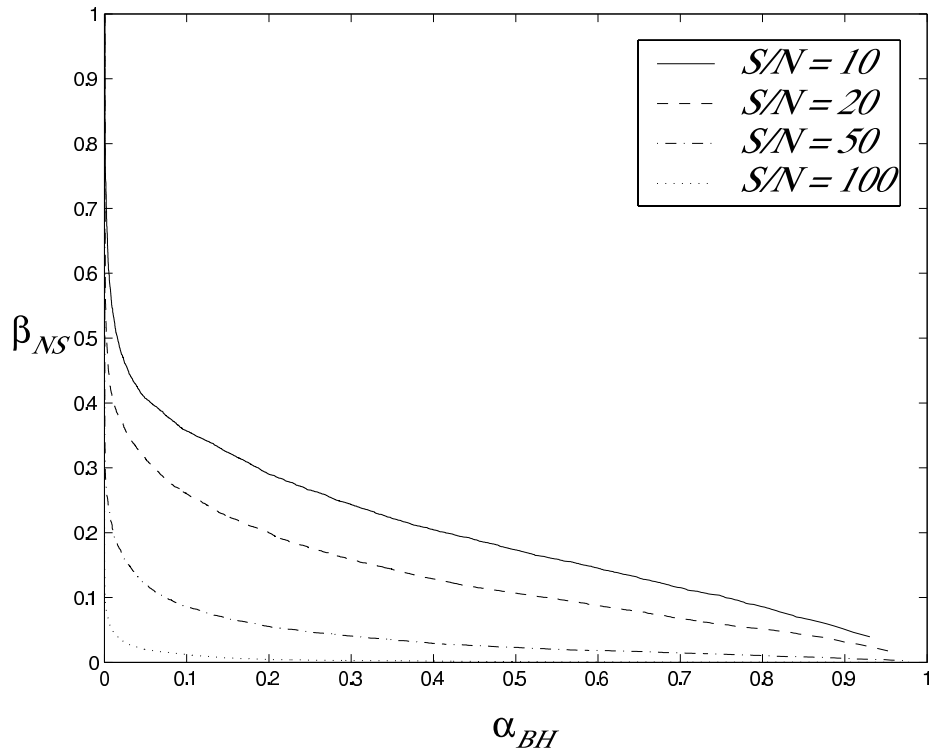}
  \caption{False dismissal probability as a function of  
  false alarm probability $\beta(\alpha)$. The false dismissal 
  probability depends on the non-black hole QNM spectrum, which we have
  taken to have the same ratio of frequencies and relationship between frequencies
  and damping times as neutron star w-modes.}\label{beta}
  \end{center}
\end{figure}

The false dismissal probability depends on the detailed character of the source,
which is not a black hole.  Strong gravitational wave sources are compact, with 
radius $R$ on order their mass $GM/c^2$ and oscillations periods of order $GM/c$.
At the frequencies where LISA will have its greatest sensitivity --- $10^{-2}$--$10^{-4}$~Hz,
corresponding to masses of order $10^6$--$10^8$~$\mbox{M}_\odot$
--- we know of no compact sources that are not black holes. For the purpose of illustration 
and to give a sense of the ability of the test
described here to ``discover'' new physics, we suppose a population of sources
whose frequencies and damping times share the same relationship as certain 
neutron star w-modes calculated in \cite{andersson95}. Referring to 
\cite[table 1, col.\ 1, lines 3, 5]{andersson95}
we consider observations consisting of two modes
\begin{eqnarray}
M\omega_1 &=& 0.471 +  0.056i,\nonumber\\
M\omega_2 &=& 0.654 +  0.164i,
\end{eqnarray}
where $M$ is drawn from the distribution given in equation (\ref{eq:distrM}).  
In exactly the same way that we used simulations in section \ref{sec:alpha} 
to determine $\alpha$ as
a function of $p$ we calculate from these simulations $\beta$ as a function of $p$. 
Together
$\alpha(p)$ and $\beta(p)$ determine $\beta(\alpha)$, which 
we show in figure 
\ref{beta}. A measure of the effectiveness of the test is the degree to which the 
curves for different signal-to-noise fall below the $\beta =1-\alpha$ 
diagonal. (A ``test'' that randomly picked a fraction $\alpha$ of
observations as not black holes would have $\beta=1-\alpha$. Any 
``test'' that can do better than randomly choosing in this way will have 
a $\beta(\alpha)$ curve that falls below this diagonal.)
As expected the test also does better with stronger signals. Consider a false 
alarm threshold of 1\%. Then for observations with S/N equal to 10 
we have a better-than-40\% 
chance of distinguishing NBH sources from BH sources. This climbs to 
better-than-90\% chance for observations with 
S/N 100. 

\section{Potential for application}
\label{sec:disc}

We have shown that, given at least two QNM signals, from the same source and 
with sufficiently large signal-to-noise,
we can cleanly distinguish black holes from other astrophysical sources. Following
\cite{finn92a} and the discussion above we assume that an amplitude-squared 
signal-to-noise ratio of $100$ for the weaker mode is sufficiently large.
In this
section we show that LISA detector observations should provide many examples
of such observations, permitting the use of this test to uniquely identify the existence
of general relativistic black holes.  

Focus attention on each individual QNM. The signal strength,
characterized by the signal-to-noise ratio at the detector, depends on
the energy radiated in the mode, the radiation pattern associated with
the mode, and the relative orientation of the detector and the
source. We can average over these latter angles to obtain the
mean-square signal-to-noise associated with the $n{\ell}m$ mode as a
function of the mode energy \cite[eq. 2.30]{flanagan98a}:
\begin{equation}
{\langle \rho^{2}\rangle} = {2(1+z)^{2}\over 5\pi^{2}D(z)^{2}}
{\int _{0}^{\infty}} df {1\over f^{2}S_{n}(f)} {dE_e\over
{df_e}}[(1+z)f] \label{eqn:flanagan1}
\end{equation}
where $z$ and $D(z)$ are, respectively, the redshift and the
luminosity distance to the source.

For the QNM, we take
\begin{eqnarray}
h_{n \ell m} & = &{A}_{n{\ell}m}
\exp{\left( -\frac{\pi f_{n{\ell}m} t}{Q_{n{\ell}m}}\right)}
\sin{\left( 2\pi f_{n{\ell}m} t \right)}
\end{eqnarray}
where $Q_{n{\ell}m} \equiv \pi f_{n{\ell}m} \tau_{n{\ell}m}$.

Note that $Q_{n{\ell}m}$, which is an observable property of a
QNM, is independent of source redshift, while the observed
$f_{n{\ell}m}$ and $\tau_{n{\ell}m}$ depend on redshift.

The ringdown energy spectrum of the $n{\ell}m$ mode is taken from eq
(3.18) of \cite{flanagan98a}:
\begin{eqnarray}
\frac{dE_e}{df_e} & = &
\frac{\epsilon_{n{\ell}m}}{F_{n{\ell}m}}\frac{Q_{n{\ell}m}}{\left(4Q_{n{\ell}m}^2
+ 1\right)}\frac{M^2 f^2}{\pi^3\tau^2} \left[ \frac{1}{[(f -
f_{n{\ell}m})^{2} + (2\pi\tau)^{-2}]^{2}} \right. \nonumber \\
&& \qquad{}+\left. \frac{1}{[(f + f_{n{\ell}m})^{2} +
(2\pi\tau)^{-2}]^{2}} \right]\label{eqn:flanagan2}
\end{eqnarray}
where the mode amplitude $A_{n\ell m}$ has been replaced with the
fraction $\epsilon_{n\ell m}$ of the mass radiated in that mode, defined by
\begin{equation}
\epsilon_{n{\ell}m} := \frac{1}{M} \int _{0}^{\infty} {dE \over
df} df.
\end{equation}
Using this spectrum in the
formula above, and assuming $S_n(f)$ is constant over the signal band
we can integrate over frequencies and invert the result to obtain an
approximate range over which we can observe a mode $n{\ell}m$ with
signal-to-noise $ > \rho_{n{\ell}m}^2$:

\begin{equation}
D(z)^{2} = {8\over 5\pi^2} {Q_{n{\ell}m}^{2}\over
4Q_{n{\ell}m}^{2} + 1} {(1+z)^{3}M^{3}\over F_{n{\ell}m}^{2}}
{\epsilon_{n{\ell}m} \over S_{n} {\rho_{n{\ell}m}^{2}}}{G^{3}\over
c^{7}}\, .\label{eqn:range_res}
\end{equation}

Given a threshold $\rho_{n{\ell}m}^2$, black holes radiating a
fraction $\epsilon_{n{\ell}m}$ of their rest energy in mode $n{\ell}m$
are observable with a redshift $z$ satisfying eq
(\ref{eqn:range_res}). To use this relation, we must make some
reasonable assumptions about the observed modes:

Numerical simulations (eg., \cite{brandt95}) suggest that energy
emitted in QNMs during ringdown can be as high as $3\%$ of the
rest-mass energy of the hole. For equal-mass black hole mergers, the
simulations suggest that the 
$\ell = 2$ modes will be by far the strongest, with total emitted
energies greater than $\ell = 4$ modes by as much as three orders of
magnitude (see \cite{anninos95}). Here we assume that the weaker mode 
of a QNM pair carries away a fraction $10^{-5}$ of
the black hole mass.

For LISA, the noise power spectral density is expected to be least in
the $10^{-3} \leq f \leq 10^{-2}$ Hz band, where it is estimated to be
$5 \times 10^{-45}$ Hz$^{-1}$. We take a frequency $f = 10^{-2}$ Hz here,
with this minimum noise power spectral density as $S_{n}$.

The relationship between luminosity distance and redshift we take to
be given by eqs (23,25) of \cite{carroll92}. This depends on
cosmological parameters $H_0$ (the Hubble parameter), $\Omega_M$,
$\Omega_{\Lambda}$ and $\Omega_k$; we use values from a recent review
\cite{wmap03a}.

Finally we can estimate the redshift within which we can expect that
supermassive black hole mergers will be visible. In addition to the
above assumptions, we require a signal-to-noise in the weaker mode of at least
$\rho ^2 = 100$. Then, assuming both modes are visible in the LISA
band, LISA will observe the ``average'' merger within a redshift of
$\sim 475$ (for extremal-spin Kerr) or $\sim 320$ (for zero-spin
Kerr), that is, the entire observable universe.

The rate of merger of black holes of the appropriate mass depends on
redshift, both because of evolution, and because the observed
frequency scales as $[(1+z)M]^{-1}$. The estimates of the event rate
out to, say, $z=1$, then range from $0.1--100 /yr$ (see
\cite{haehnelt98a},\cite{cutler02a} for recent reviews).

We conclude that, if any mergers are observable, there is good
reason to believe they will result in strong QNM excitation and
several modes will be observable and separately identifiable.

\section{Conclusion}
\label{sec:conc}

We have described a qualitatively new test of the existence of 
general relativistic black holes, based on the gravitational radiation they emit
when they are formed or when they are impulsively excited, e.g., through a merger
event. 
Radiation from an impulsively excited
black hole, such as might arise in the course of a non-spherical black 
hole formation 
event or 
the coalescence of
a black hole with another black hole or compact object, has a component
that consists of a sum of damped sinusoids. This signature is, characteristic
of the radiation from any impulsively excited, damped source. For any given 
mode, the
scale of the frequency and damping time measures the black hole mass
and angular momentum. Similarly, the relationship of the different modes to 
each other
--- i.e., the spectrum --- is unique to black holes. 
We have described here how this relationship can be used to test the
proposition
that observed gravitational waves, characteristic of an impulsively excited,
damped source, in fact originate from a general relativistic black hole. 
Such a test can be characterized in at least two different ways: as a definitive
``proof'' that a black hole has  been observed, or as a test of the so-called
``no-hair'' theorem of general relativity. 

To demonstrate the effectiveness of this test we have evaluated numerically
the probability that the test will mistakenly fail to identify an actual black hole.
By introducing a hypothetical 
gravitational wave source whose characteristic frequencies and damping times 
are similar to those of neutron star w-mode \cite{andersson95} we have also
evaluated numerically the probability that the test will incorrectly identify w-mode 
oscillations of a neutron star, or any object whose spectrum is similar to a 
black hole. 
Together these results demonstrate that 
for sources with the signal-to-noise expected of, for example, massive black 
hole coalescences detected by LISA, the test proposed here can cleanly 
discriminate black hole sources. Finally, we have shown that LISA can be
expected to observe signals of this kind and strength throughout the universe
with a rate that may be as great as 100/y.  

This method can be used to measure
mass and angular momentum of a black hole. Using gravitational
waves to measure mass and angular momentum is an idea that has
been around for some time \cite{eche89,finn92a}. In these previous works
it was assumed that the mode observed was of a known order (e.g., the mode
with the longest damping time, or the lowest order, etc.). With the observation 
of two or more modes the requirement that a single mass and angular momentum
explain the complete set likely permits the mass and angular momentum to be 
determined uniquely. 

The field of gravitational-wave detection is new.  
The current generation of ground- and space-based
gravitational-wave detectors is opening a new frontier of physics:
\emph{gravitational-wave phenomenology,} or  the use of 
gravitational wave observations to learn about the physics of gravitational-wave
sources and gravity itself. 
We are only just
beginning learn how to exploit the opportunities it is creating for us.
As gravitational-wave observations
mature, we can expect more and greater recognition of their utility as
probes of the character of relativistic gravity.  The opening of this
new frontier promises to be an
exciting and revealing one for the physics of gravity.

\ack 
We are grateful to Soumya Mohanty and Curt Cutler for 
helpful discussions.  We
acknowledge support through the Center for Gravitational Wave Physics,
which is funded by the NSF under cooperative agreement PHY 01-14375,
NSF awards PHY 00-90091 and PHY 00-99559, the Eberly 
research funds of Penn State, and the Albert-Einstein Institut.

\end{document}